\documentclass[aps,prl]{revtex4-1}
\usepackage{graphicx}
\begin{document}
\title{Dislocations in magnetohydrodynamic waves in a stellar atmosphere}
%\draft
\author{A. L\'{o}pez Ariste}
\affiliation{THEMIS  - CNRS UPS 853. C/ V\'{\i}a L\'{a}ctea s/n. 38200 La Laguna. Spain}
\author{M. Collados}
\author{E. Khomenko}
\affiliation{Instituto de Astrof\'{\i}sica de Canarias. C/ V\'{\i}a L\'{a}ctea s/n. 38200 La Laguna. Spain}
\affiliation{Departamento de Astrof\'{\i}sica, Universidad de La Laguna, 38205 La Laguna, Tenerife, Spain}

\date{\today}
\begin{abstract}
We describe the presence of wavefront dislocations in magnetohydrodynamic waves in stratified stellar atmospheres. Scalar dislocations such as edges and vortices 
can appear in Alfv\'en waves, as well as in general magneto-acoustic waves. We detect those dislocations in observations of magnetohydrodynamic waves in sunspots 
in the solar chromosphere. Through the measured charge of all the dislocations observed, we can give for the first time estimates of the modal contribution in the waves propagating
along magnetic fields in  solar sunspots.
\end{abstract}
\maketitle

\begin{figure}
 \includegraphics[width=6cm]{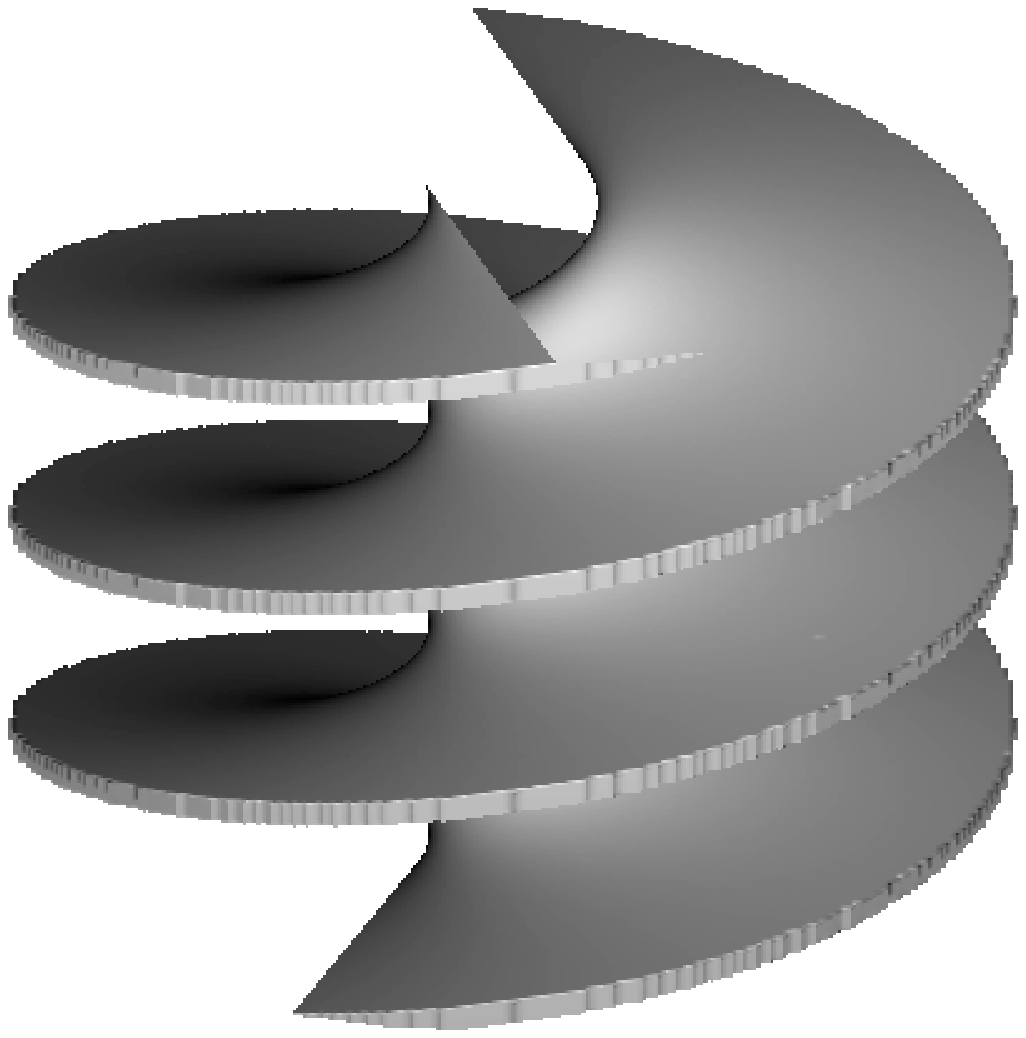} \includegraphics[width=6cm]{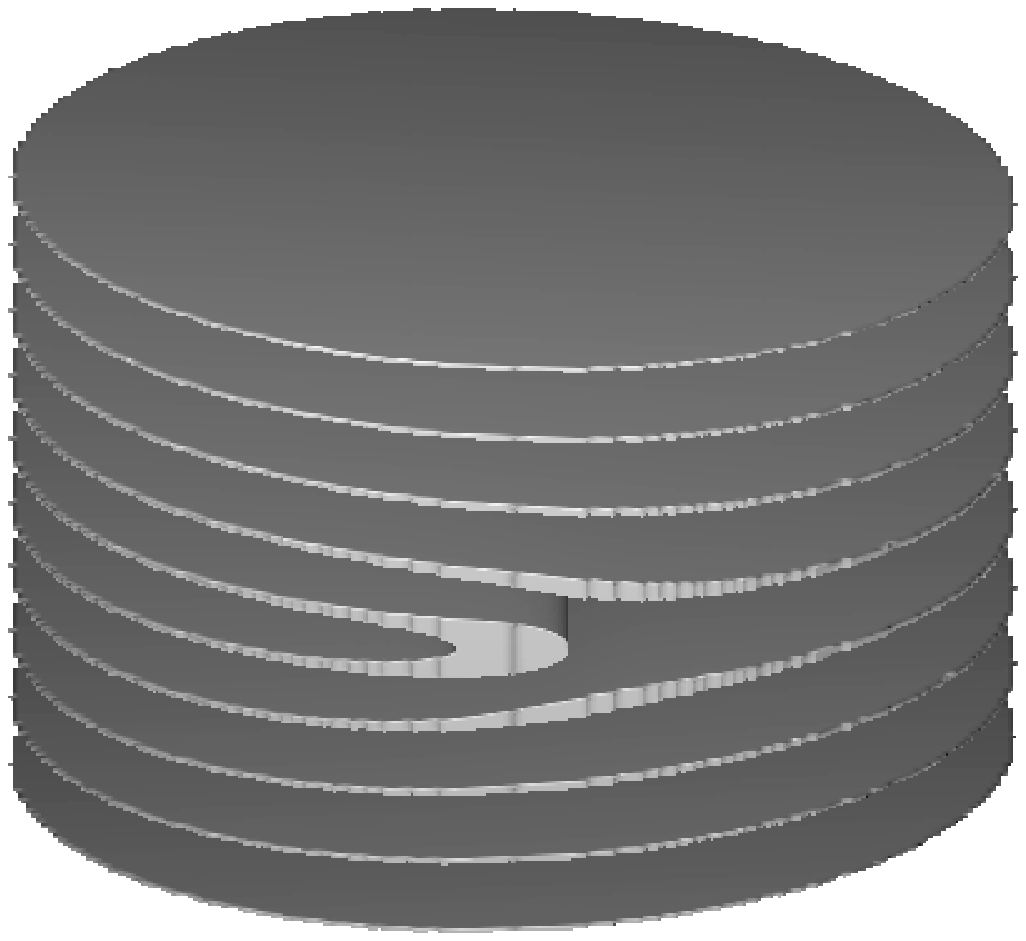}%
 \caption{Depictions of the wavefronts of a wave propagating upwards and carrying a vortex dislocation (left) and an edge dislocation(right). In the vortex case
 the phase is singular along the axis of the helix, while in the edge it is singular along the folding axes of the iso-phase (modulo $2\pi$) surface.}\label{3d}
 \end{figure}

{Wavefront dislocations are singularities of the phase of the wave \cite{nye_dislocations_1974}. At any given time 
and space position we can always describe the wave as $Ae^{i\chi}$ where $A$, the amplitude, and $\chi$, the phase, are real quantities. However the wave equations, 
as demonstrated by  \cite{nye_dislocations_1974}, allow the possibility that, at certain positions and times, the phase $\chi$ is singular and not defined. To preserve
the properties of the wave, the amplitude at that point and time has to be identically zero. When such a dislocation appears in a wave, the topology of the wavefront 
is analogous to that of dislocations in a crystal \cite{nabarro_theory_1967}, hence the use of the same word to refer to these wave features. Following this analogy, the two main classes 
of crystal dislocations are also
found in wavefront dislocations, and the same names - vortices and edges - are kept to refer to them. Figure \ref{3d} shows examples of wavefronts in an upward 
propagating wave
for the  most simple case of respectively a vortex and an edge dislocation. Dislocation-carrying waves have interesting properties. An 
example of the changing properties of these waves is the integral $$\oint _C \vec{\nabla} \chi \cdot d\vec{r}$$ over a closed circuit.
For any finite combination of plane waves, the phase is well defined at
every point enclosed by the circuit $C$ and Stokes's theorem indicates that the integral is zero. But in the presence of a singularity, Stokes's theorem fails 
and the integral above becomes $n\pi$, where
$n$ is an integer that is sometimes called the \textit{charge} of the dislocation. This charge associated with any dislocation is actually a conserved quantity of the wave, 
and once a wave is found to carry  a given charge it should conserve it, unless the wave is dissipated or collides with another wave with a different
charge \cite{lenzini_optical_2011}.
The conservation of the charge of the wave can be seen to survive even when a change in the medium converts it into a shock wave \cite{brunet_experimental_2009}. 
Dislocations are therefore highly resilient features of a wave.}

Wave dislocations as introduced by \cite{nye_dislocations_1974} have been studied on surface water and analogous quantum mechanical waves\citep{berry_wavefront_1980} 
and sound  waves\cite{nye_dislocations_1974,brunet_experimental_2009}. Interest in them
has grown recently since the discovery of vortex dislocations in light. {But dislocations are general features of waves and one would expect them to be
found also in magnetohydrodynamic waves.}
We intend here to study their existence and properties in magnetohydrodynamic 
waves travelling through stellar atmospheres \cite{ferraro_hydromagnetic_1958,goossens_resonant_2011}. Although many of the properties of the dislocations found are 
probably general, we are going to limit this
study to the case of waves in an 
isothermal non-stratified plasma where the magnetohydrodynamic approximation applies, as is the case in many stellar atmospheres, particularly in the solar 
photosphere, with the only addition of stratification. Observation of magnetohydrodynamic waves in the solar photosphere already provides 
examples of dislocations that can now be correctly interpreted and analysed.

{Plane waves are not an appropriate family of solutions for describing dislocations, so we must rely on 
other families of solutions of the wave equation. Some of them actually carry dislocations in each component. One example  appears 
when solving 
the wave equation for cylindrical boundary conditions when the Bessel function times and azimuthal phase component appear as the appropriate family of solutions 
for the wave.}

{After this general reminder of the concept of wave dislocation,} our approach is going to be a straightforward one:
We address a differential equation for magnetohydrodynamic waves in a static stellar atmosphere and we will test whether
waves carrying dislocations are solutions to this equation.
The equation we chose to address is Eq. 4.14 from  Priest \cite{priest_solar_1982}:
\begin{equation}
 \frac{\partial^2 \vec{v}}{\partial t^2} = c_S^2\vec{\nabla}(\vec{\nabla}\cdot \vec{v})+
\left[\vec{\nabla}\times(\vec{\nabla}\times(\vec{v}\times\vec{B}_0))\right]\times\frac{\vec{B}_0}{\mu \rho_0},
\label{eq}
\end{equation}
\textbf{where $\vec{v}$ is the velocity perturbation introduced by the wave into the plasma}, $\vec{B}_0$ 
 \textbf{is the background } magnetic field, assumed constant 
\textbf{in space and time}, $c_S$ the speed of sound in the medium,
 $\rho_0$  the density \textbf{and $\mu$ the vacuum permeability}. This equation describes magnetohydrodynamic waves in a homogeneous, isothermal medium. Although it is 
not the most  general equation for waves in a magnetized atmosphere, it has the advantage of relative simplicity while still capturing many
of the physical phenomena related to those waves. 
We aim to find solutions that are valid  in at least a small enough region around the phase singularity which
defines the dislocation, though perhaps not in the total volume.  By requiring such test solutions to be valid in a finite region rather than 
in the full domain we avoid the problems
related to the total energy carried by such waves and can also afford to assume many quantities, such as the magnetic field being constant over the region
of validity. Again, this simplifies our computations but it also allows us to justify such an approximation, since we can always assume that, independently
of how the magnetic field varies in the full atmosphere, there is always a small enough region where it can be approximated by a constant value.

As it is usual when considering such a wave equation, we distinguish between two cases. In the first one we study Alfv\'en waves, which are \textbf{incompressible 
transverse waves propagating at the Alfv\'en speed}. In the second case we address  magneto-acoustic waves that 
have a longitudinal wave component.
 One can further distinguish between
fast and slow waves among these magneto-acoustic waves by their phase velocity, but we will not enter into such distinctions in this article. We will eventually
restrict ourselves to the particular case where the direction of propagation of the wave is parallel to the magnetic field, a case in which observations 
of waves in the solar atmosphere are easier to interpret. We will refer to that direction as \textit{vertical} henceforth.

In the case of both Alfv\'en and magneto-acoustic waves, we try solutions carrying either an edge or a vortex dislocation, {as those shown in Figure \ref{3d}. A
simple mathematical description of these dislocations for waves solution of the scalar wave equation \textbf{propagating in a given direction $z$ ($x$ and $y$ being 
the coordinates in the transverse plane)} has been given by \cite{Berry105}}:
\begin{itemize}
 \item Edge dislocation: $v=A(kx+i\beta k\zeta)e^{ik\zeta}$, where $A$ and $\beta$ are real constants, $k$ is the wavenumber, assumed to be constant, and
$\zeta=z-ct$, where $c$ is the speed of the wave, either the Alfv\'en speed or the speed of sound.
\item Vortices: $v=A(kx \pm iky)^m e^{ik\zeta}$, where $m$ is an integer, usually called the topological charge of the dislocation. 
\end{itemize}

{It is important to stress that the} above expressions define scalar, not vector, waves. On the other hand, Eq.\ \ref{eq} is a wave equation for a vector quantity, the velocity
$\vec{v}$. We therefore need to define how the scalar solution is related to the vector $\vec{v}$. 
In the case of Alfv\'en waves this is 
straightforward since, due to the definition
of those waves, they are actually polarized waves for which the scalar solution 
can be used for the components of the velocity in  the plane transverse to the
direction of propagation. The case of magneto-acoustic 
waves, on the other hand, requires solving for two components of the velocity vector. Because of the comparison with real observations of waves in 
the solar atmosphere {that will be made below,} we impose the scalar solution with a dislocation to the vertical component of the velocity 
perturbation and check for consistency of solutions in all three components, {with positive results. }

Both the Alfv\'en and magneto-acoustic cases accept edges and vortices as solutions in the conditions specified, and this is the first 
result from this study. Although we have introduced several simplifications to allow  easy interpretion of real observations in the rest of the paper, 
{ dislocations become more common only as the wave equation becomes more complex. An example of this is electromagnetic waves, for which scalar dislocations
analogous to those studied in this paper are just a particular case of the more general vector dislocations that appear when one considers the polarization
of light \cite{nye_lines_1983}.
Therefore,} we do not anticipate any serious concern regarding dislocations being a solution under less restrictive conditions than those described here.

To stress this point, it is interesting to consider the situation of a magnetic flux tube with diameter comparable
to the wavelength. These cases have been amply studied in the literature because of their obvious importance in the study of waves in magnetized
stellar atmospheres \cite{edwin_wave_1983,ruderman_nonlinear_2010,cally_leaky_1986}. The many investigations of those particular conditions all use 
the previous equation or analogous ones {to describe the waves} and
impose continuity  conditions on the physical variables (such as gas pressure and density) at the walls of the flux tube. The cylindrical constraint imposed by
such continuity
conditions and the geometry of the flux tube suggest as solutions for the longitudinal component of the velocity waves propagating along the axis of the cylinder 
whose amplitude is written in terms of Bessel functions for the radial dependence times an azimuthal 
dependence in $e^{im\theta}$, where $\theta$ is the
azimuthal coordinate in the transverse plane and $m$ is the order of the Bessel function. Those solutions naturally carry a vortex dislocation at $r=0$, in the
axis of the flux tube, with a topological charge equal to the order $m$ of the Bessel function and the azimuthal dependence. {This is one of those families
of solutions to the wave equation that, unlike plane waves, carries dislocations in every component.} 
It has been 
customary\citep{edwin_wave_1983,bogdan_normal_1989}
to call solutions with $m=0$ and no dislocations \textit{sausage modes}, while solutions with $m=1$ carrying a vortex dislocation are 
referred to as \textit{kink modes}.

\begin{figure}
 \includegraphics[width=15cm]{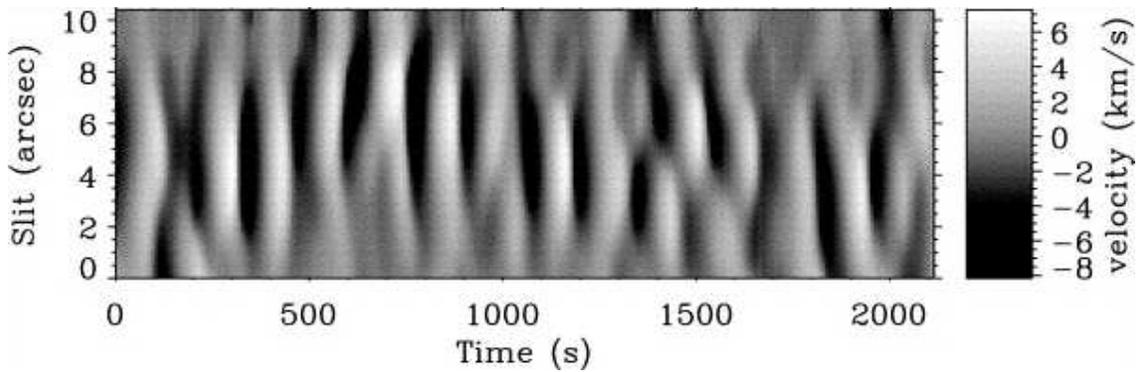}%
 \caption{Velocity in the line of sight measured as the Doppler shift of an atomic spectral line measured in the solar chromosphere across a sunspot over 2000 
 seconds. Magneto-acoustic waves are seen with typical periods of 3 minutes.Observations by \citep{centeno_spectropolarimetric_2006}}\label{figure}
 \end{figure}

We now turn our attention to the direct observation of dislocations in magnetohydrodynamic waves in stellar atmospheres. 
 Figure \ref{figure} shows a typical observation of magnetohydrodynamic waves in a sunspot\citep{centeno_spectropolarimetric_2006}. It shows the Doppler shift (given in terms of velocity 
 along the line of sight)  of a spectral line forming in the solar chromosphere as a function of time and along a line (the slit of the spectrograph)
 passing through a sunspot. A wave with a typical period of around 3 minutes is seen over almost  the whole width of the sunspot. Since the 
 observation was made when the sunspot 
 was near the centre of the solar disc, the line of sight coincides with the local vertical and the wave is made of velocity oscillations along this 
 direction. We have no information about possible oscillations in the other components of the velocity vector. However, we know that the magnetic
 field is predominantly vertical in the centre of the sunspot, and that, in consequence, the velocity oscillations seen in the figure are oscillations
 along the magnetic field vector. These arguments suggest that the figure shows a magneto-acoustic wave propagating with the speed of sound and falling
 into the cases we studied in detail above.
 
 During the more than 2000 seconds of the observation we observe around 14 periods, about half of them presenting dislocations of one kind or another.
 Perhaps the clearest cases are the forks around second 1700  and also the gliding dislocation occurring between 1300 and 1600 seconds, a mixture of
 edge and vortex that we should mathematically describe below. The 
 identification
 of those patterns as dislocations follows from the identification of the amplitude modulation as due to the phase of the wave, which is the 
 straightforward interpretation of the signal.

\begin{figure}
 \includegraphics[width=12cm]{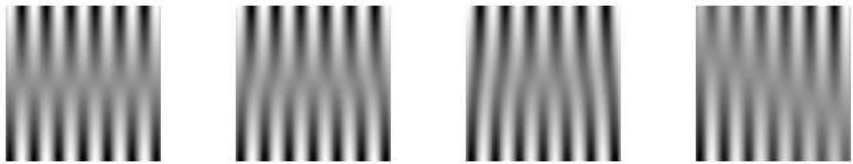}% Use ~/DISLOCACIONES/disloc3.pro to create it
 \caption{Four examples of dislocations as would be observed by an instrumental setup such as that used for the observations of waves in solar sunspots in Figure~\ref{figure}. 
 In the for images, time runs horizontally and the vertical axis is the spatial direction along a simulated slit accross the sunspot. Starting on the left the figures
 show a vortex ($\delta=\frac{\pi}{2}$ and $\beta=0$), two edges ($\delta=\frac{\pi}{2}$ and $\beta=0.4$ and 1) and one gliding dislocation ($\delta=0$ and $\beta=0.9$).
 }\label{figure2}
 \end{figure}

 After the above theoretical analysis, we know that both edge and vortex dislocations are allowed for the waves in Fig.\ref{figure} that are assumed to 
 be the longitudinal component of 
 a magneto-acoustic wave propagating parallel to the magnetic field of the sunspot. To help us characterize the observed dislocations we will propose a wave solution
 in cylindrical coordinates of the form
 $$v_z=A(kr^me^{im\theta}+\beta e^{i\delta} k \zeta)e^{ik\zeta},$$
 where all the symbols retain the meaning given above, $z$ represents the vertical direction, and we have introduced a phase $\delta$ and a relative amplitude ratio $\beta$ between the two contributions
 to the wave amplitude. {This  proposed solution for $v_z$ allows us, by varying these phase and relative amplitudes, to produce all sorts of dislocations: a 
 vortex  appears whenever $\beta=0$ with different charges $m$, and a pure edge when $\beta\ne 0$ and $\delta =\pi/2$. The gliding dislocation mentioned above 
 appears easily by putting $\delta =0$.
  Figure~\ref{figure2} shows four 
 examples of simulated observations like those in Fig.~\ref{figure} with different combinations of $\beta$ and $\delta$. The first example on the left is a vortex
 with $\beta=0$ characterized by its constancy in time, in abscissas. The second and third examples are pure edges with $\delta =\pi/2$ and two values of $\beta$. 
These three cases are 
 actually cuts of the 3-dimensional wavefronts illustrated in Fig.~\ref{3d} along a vertical plane. The last example on the right is a gliding dislocation, made by
 setting $\delta=0$. The dislocation is seen to move spatially (along the ordinates) with a certain speed with respect to the wave phase speed set by the value of $\beta$: 
 the dislocation is \textit{surfing} or \textit{gliding} the wave.}

 {For completeness, we also compute the transverse velocity components corresponding to the solution proposed for $v_z$. Eq.\ref{eq} can be rewritten 
 as a coupled system of three equations: }
 $$  \frac{\partial^2 v_z}{\partial t^2} = c_S^2\frac{\partial}{\partial  z} \frac{1}{r}\left[\frac{\partial}{\partial r} (rv_r)+\frac{\partial}{\partial \theta} v_{\theta}\right]  + c_S^2\frac{\partial^2}{\partial  z^2 } v_z$$
 
 $$ \frac{\partial^2 v_r}{\partial t^2} = (c_S^2+v_A^2)\frac{\partial}{\partial r} \frac{1}{r}\left[ \frac{\partial}{\partial r} (rv_r) + \frac{\partial}{\partial \theta} (v_\theta)\right] +    c_S^2\frac{\partial^2}{\partial r\partial z } v_z + v_A^2\frac{\partial^2 v_r}{\partial z^2} $$
 $$  \frac{\partial^2 v_\theta}{\partial t^2} =(c_S^2+v_A^2)\frac{\partial}{\partial \theta} \frac{1}{r}\left[ \frac{\partial}{\partial r} (rv_r) + \frac{\partial}{\partial \theta} (v_\theta)\right]+c_S^2\frac{1}{r}\frac{\partial^2}{\partial \theta \partial z } v_z + v_A^2\frac{\partial^2 v_\theta}{\partial z^2}$$ 
 \textbf{with $v_A=\frac{B_0}{\sqrt{\mu \rho_0}}$ the Alfv\'en speed.}
 
%% $$ \frac{\partial^2 v_x}{\partial t^2} = v_A^2  \frac{\partial^2 v_x}{\partial z^2}+c_S^2 A i k^2  \frac{\partial }{\partial x}\left( J_m e^{im\theta}\right)e^{ik\zeta},$$
We propose the solution for $v_z$ given above and, if we assume that the transverse divergence of the horizontal velocity is zero,
$$\vec{\nabla}_T \cdot \vec{v}_T = \frac{1}{r}\left[ \frac{\partial}{\partial r} (rv_r) + \frac{\partial}{\partial \theta} (v_\theta)\right] =0,$$
 a condition automatically verified by Alfv\'en waves, but not necessarily for magneto-acoustic ones, it is stragihtforward to show that 
$$ v_r=\frac{c_S^2}{v_A^2-c_S^2}i \frac{A}{k} m r^{m-1}e^{im\theta}e^{ik\zeta}$$
$$ v_{\theta}=-\frac{c_S^2}{v_A^2-c_S^2} \frac{A}{k} m r^{m-1}e^{im\theta}e^{ik\zeta}= iv_r$$
as long as $v_A\ne c_S$. In the general magneto-acoustic case in which the transverse divergence $\vec{\nabla}_T \cdot \vec{v}_T $ is not zero, expressions are more elaborate but it is
relatively straightforward to give a solution for this transverse divergence as
$$\vec{\nabla}_T \cdot  \vec{v}_T =A(k r^{m}  e^{im\theta}+\beta e^{i\delta} k \zeta )e^{ik\zeta} $$
of the same form as $v_z$.

{Using the above solution for $v_z$}, we can
 reproduce all the observed dislocations in  Figure~\ref{figure}. Most of them appear to be of the case with a $\delta=\pi/2$ phase difference between 
 both components, prompting the question about what in the solar atmosphere favours such case. But we can also see cases with $\delta=0$, the most conspicuous 
 of which being the gliding dislocation referred to above. In all cases, including other observations studied and not shown here, there appears to be no contribution
 from waves with $m>1$, and the only excited waves are those with $m=1$ and $m=0$.
 
 The use of the azimuthal number $m$ and the radial functions can be misleading. Although a sunspot can be considered as a large magnetic flux tube, its scale
 is much larger than the wavelength and therefore more often than not the wave is formed and propagated with no influence from the sunspot's cylindrical boundaries.
Because of this we have avoided the use of the words \textit{kink} and \textit{sausage} to refer to the $m=1$ and $m=0$ modes identified in the waves thanks to 
the observed 
 dislocations. The above solution for $v_z$ is a solution of the wave equation with no reference to particular geometries and/or boundary conditions, and the 
 observations suggest that those waves do propagate in the real solar atmosphere independently of  any cylindrical geometry. So it is legitimate to write such a solution
 and to characterize those waves by their $m$ number.
 
 A second interesting point to be stressed is that the waves we identify through the observed dislocations are the coherent addition of modes with
 $m=0$ and $m=1$ with definite phase relationships. That is, these are not waves which can be formed by the incoherent addition of two physically unrelated waves. 
 The two modes are excited by the same physical mechanism and evolve in a coherent manner in their propagation through the solar atmosphere.

 \section{Summary and conclusions}
 We have shown that magnetohydrodynamic waves such as those found in stellar atmospheres present phase singularities or dislocations, in a similar way as 
 previously found for sound and electromagnetic waves. In the 
 simplified case of an isothermal atmosphere both Alfv\'en and magneto-acoustic waves can carry scalar dislocations of either the edge or vortex kind.
 We have re-examined observations of the longitudinal component of the velocity in magneto-acoustic waves in
 sunspots and found the signature of dislocations. We  identify these predominantly as edges and gliding edges. Such dislocations can easily 
 be described as the coherent addition of two modes, one without dislocations characterized by an azimuthal number $m=0$ and the other carrying a vortex
 dislocation ($m=1$). The phase and amplitude relations between both modes describe all observed dislocations with a predominant phase difference of
 $\pi/2$. The observations show that at least half
 of the observed periods carry dislocations, and we conclude that the excitation of those two modes with the fixed phase relation between them should be a 
 favoured behaviour of the wave excitation mechanism.

 The other half of  the observed periods either carry no dislocation
 or it was missed by the position of the slit or the height of formation of the spectral line observed.  We can therefore only set a lower bound
 of 50\% on the number of wave periods carrying a dislocation {of the type $m=1$. Although it is known from theory that modes with $m=1$ and $m=0$ are 
 the most likely to be excited. This is the first time, to our knowledge, that such a result has been verified observationally.}
 
 Other than the presence of dislocations in MHD waves, we wish to stress as  a result of this work the importance of the observation of dislocations
 in observed waves, as in the Sun, to determine the nature and modal properties of those waves. An example of such diagnosing capabilities has been
 given here by determining from observations that at least 50\% of the wave periods correspond to a wave with an $m=1$ component. 
 
 {We have not addressed what the mechanisms might be that generate such dislocations. In many cases, a cylindrical boundary is enough to generate a vortex 
 dislocation. Flux tubes in the solar atmosphere must therefore be prone to show dislocations in the waves that propagate along them. It has also been found
 that a wave propagating through a fluid with non-zero vorticity adquires an edge dislocation \citep{berry_wavefront_1980}. It is tempting to think that edge 
 dislocations in solar waves can also be a tracer of regions with high vorticity along the wave path and use them to determine the vorticity in the solar 
 atmosphere. Finally, it  is also known }that vortex dislocations carry torque and that this torque can be transferred to an absorbing medium
 \cite{allen_orbital_1992,simpson_optical_1996,demore_mechanical_2012,volke-sepulveda_transfer_2008}.
   {In the solar atmosphere the torque that might be transferred to the solar coronal plasma by upward propagating waves could have interesting implications 
 in the dynamics of the corona.}

% body of paper here

% now the references. delete or change fake bibitem. delete next three
%   lines and directly read in your .bbl file if you use bibtex.
%\begin{references}
%\bibitem{tag} Fake bibitem.
%\end{references}
%\bibliography{/home/arturo/TeX/0biblio/aamnem99,/home/arturo/TeX/0biblio/biblio,/home/arturo/TeX/0biblio/articulos,/home/arturo/Dropbox/art44}

\begin{thebibliography}{10}%
\makeatletter
\providecommand \@ifxundefined [1]{%
 \ifx #1\undefined \expandafter \@firstoftwo
 \else \expandafter \@secondoftwo
\fi
}%
\providecommand \@ifnum [1]{%
 \ifnum #1\expandafter \@firstoftwo
 \else \expandafter \@secondoftwo
\fi
}%
\providecommand \enquote [1]{``#1''}%
\providecommand \bibnamefont  [1]{#1}%
\providecommand \bibfnamefont [1]{#1}%
\providecommand \citenamefont [1]{#1}%
\providecommand\href[0]{\@sanitize\@href}%
\providecommand\@href[1]{\endgroup\@@startlink{#1}\endgroup\@@href}%
\providecommand\@@href[1]{#1\@@endlink}%
\providecommand \@sanitize [0]{\begingroup\catcode`\&12\catcode`\#12\relax}%
\@ifxundefined \pdfoutput {\@firstoftwo}{%
 \@ifnum{\z@=\pdfoutput}{\@firstoftwo}{\@secondoftwo}%
}{%
 \providecommand\@@startlink[1]{\leavevmode}%
 \providecommand\@@endlink[0]{}%
}{%
 \providecommand\@@startlink[1]{%
  \leavevmode
  \pdfstartlink
   attr{/Border[0 0 1 ]/H/I/C[0 1 1]}%
   user{/Subtype/Link/A<</Type/Action/S/URI/URI(#1)>>}%
  \relax
 }%
 \providecommand\@@endlink[0]{\pdfendlink}%
}%
\providecommand \url  [0]{\begingroup\@sanitize \@url }%
\providecommand \@url [1]{\endgroup\@href {#1}{\urlprefix}}%
\providecommand \urlprefix [0]{URL }%
\providecommand \Eprint[0]{\href }%
\@ifxundefined \urlstyle {%
  \providecommand \doi [1]{doi:\discretionary{}{}{}#1}%
}{%
  \providecommand \doi [0]{doi:\discretionary{}{}{}\begingroup
  \urlstyle{rm}\Url }%
}%
\providecommand \doibase [0]{http://dx.doi.org/}%
\providecommand \Doi[1]{\href{\doibase#1}}%
\providecommand \bibAnnote [3]{%
  \BibitemShut{#1}%
  \begin{quotation}\noindent
    \textsc{Key:}\ #2\\\textsc{Annotation:}\ #3%
  \end{quotation}%
}%
\providecommand \bibAnnoteFile [2]{%
  \IfFileExists{#2}{\bibAnnote {#1} {#2} {\input{#2}}}{}%
}%
\providecommand \typeout [0]{\immediate \write \m@ne }%
\providecommand \selectlanguage [0]{\@gobble}%
\providecommand \bibinfo [0]{\@secondoftwo}%
\providecommand \bibfield [0]{\@secondoftwo}%
\providecommand \translation [1]{[#1]}%
\providecommand \BibitemOpen[0]{}%
\providecommand \bibitemStop [0]{}%
\providecommand \bibitemNoStop [0]{.\EOS\space}%
\providecommand \EOS [0]{\spacefactor3000\relax}%
\providecommand \BibitemShut [1]{\csname bibitem#1\endcsname}%
%</preamble>
\bibitem{nye_dislocations_1974}%
  \BibitemOpen
  \bibfield{author}{%
  \bibinfo {author} {\bibfnamefont{J.~F.}\ \bibnamefont{Nye}}\ and\ \bibinfo
  {author} {\bibfnamefont{M.~V.}\ \bibnamefont{Berry}},\ }%
  \bibfield{journal}{%
  \Doi{10.1098/rspa.1974.0012}{\bibinfo {journal} {Royal Society of London
  Proceedings Series A}}\ }%
  \textbf{\bibinfo {volume} {336}},\ \bibinfo {pages} {165} (\bibinfo {month}
  {Jan.}\ \bibinfo {year} {1974}),\
  \url{http://adsabs.harvard.edu/abs/1974RSPSA.336..165N}%
  \bibAnnoteFile{NoStop}{nye_dislocations_1974}%
\bibitem{nabarro_theory_1967}%
  \BibitemOpen
  \bibfield{author}{%
  \bibinfo {author} {\bibfnamefont{F.~R.~N.}\ \bibnamefont{Nabarro}},\ }%
  \emph{\bibinfo {title} {{Theory of crystal dislocations}}}\ (\bibinfo
  {publisher} {Clarendon P.},\ \bibinfo {address} {Oxford},\ \bibinfo {year}
  {1967})\ ISBN \bibinfo {isbn} {0198512449 9780198512448}%
  \bibAnnoteFile{NoStop}{nabarro_theory_1967}%
\bibitem{lenzini_optical_2011}%
  \BibitemOpen
  \bibfield{author}{%
  \bibinfo {author} {\bibfnamefont{F.}~\bibnamefont{Lenzini}}, \bibinfo
  {author} {\bibfnamefont{S.}~\bibnamefont{Residori}}, \bibinfo {author}
  {\bibfnamefont{F.~T.}\ \bibnamefont{Arecchi}},\ and\ \bibinfo {author}
  {\bibfnamefont{U.}~\bibnamefont{Bortolozzo}},\ }%
  \bibfield{journal}{%
  \Doi{10.1103/PhysRevA.84.061801}{\bibinfo {journal} {Physical Review A}}\ }%
  \textbf{\bibinfo {volume} {84}},\ \bibinfo {pages} {061801} (\bibinfo {month}
  {Dec.}\ \bibinfo {year} {2011}),\ ISSN \bibinfo {issn} {1050-2947},\
  \url{http://adsabs.harvard.edu/abs/2011PhRvA..84f1801L}%
  \bibAnnoteFile{NoStop}{lenzini_optical_2011}%
\bibitem{brunet_experimental_2009}%
  \BibitemOpen
  \bibfield{author}{%
  \bibinfo {author} {\bibfnamefont{T.}~\bibnamefont{Brunet}}, \bibinfo {author}
  {\bibfnamefont{J.}~\bibnamefont{Thomas}}, \bibinfo {author}
  {\bibfnamefont{R.}~\bibnamefont{Marchiano}},\ and\ \bibinfo {author}
  {\bibfnamefont{F.}~\bibnamefont{Coulouvrat}},\ }%
  \bibfield{journal}{%
  \bibinfo {journal} {New Journal of Physics}\ }%
  \textbf{\bibinfo {volume} {11}},\ \bibinfo {pages} {3002} (\bibinfo {month}
  {Jan.}\ \bibinfo {year} {2009}),\
  \url{http://adsabs.harvard.edu/abs/2009NJPh...11a3002B}%
  \bibAnnoteFile{NoStop}{brunet_experimental_2009}%
\bibitem{berry_wavefront_1980}%
  \BibitemOpen
  \bibfield{author}{%
  \bibinfo {author} {\bibfnamefont{M.~V.}\ \bibnamefont{Berry}}, \bibinfo
  {author} {\bibfnamefont{R.~G.}\ \bibnamefont{Chambers}}, \bibinfo {author}
  {\bibfnamefont{M.~D.}\ \bibnamefont{Large}}, \bibinfo {author}
  {\bibfnamefont{C.}~\bibnamefont{Upstill}},\ and\ \bibinfo {author}
  {\bibfnamefont{J.~C.}\ \bibnamefont{Walmsley}},\ }%
  \bibfield{journal}{%
  \Doi{10.1088/0143-0807}{\bibinfo {journal} {European Journal of Physics}}\ }%
  \textbf{\bibinfo {volume} {1}},\ \bibinfo {pages} {154} (\bibinfo {month}
  {Jul.}\ \bibinfo {year} {1980}),\
  \url{http://adsabs.harvard.edu/abs/1980EJPh....1..154B}%
  \bibAnnoteFile{NoStop}{berry_wavefront_1980}%
\bibitem{ferraro_hydromagnetic_1958}%
  \BibitemOpen
  \bibfield{author}{%
  \bibinfo {author} {\bibfnamefont{C.~A.}\ \bibnamefont{Ferraro}}\ and\
  \bibinfo {author} {\bibfnamefont{C.}~\bibnamefont{Plumpton}},\ }%
  \bibfield{journal}{%
  \bibinfo {journal} {The Astrophysical Journal}\ }%
  \textbf{\bibinfo {volume} {127}},\ \bibinfo {pages} {459} (\bibinfo {month}
  {Mar.}\ \bibinfo {year} {1958}),\
  \url{http://adsabs.harvard.edu/abs/1958ApJ...127..459F}%
  \bibAnnoteFile{NoStop}{ferraro_hydromagnetic_1958}%
\bibitem{goossens_resonant_2011}%
  \BibitemOpen
  \bibfield{author}{%
  \bibinfo {author} {\bibfnamefont{M.}~\bibnamefont{Goossens}}, \bibinfo
  {author} {\bibfnamefont{R.}~\bibnamefont{Erd{\'e}lyi}},\ and\ \bibinfo
  {author} {\bibfnamefont{M.~S.}\ \bibnamefont{Ruderman}},\ }%
  \bibfield{journal}{%
  \Doi{10.1007/s11214-010-9702-7}{\bibinfo {journal} {Space Science Reviews}}\
  }%
  \textbf{\bibinfo {volume} {158}},\ \bibinfo {pages} {289} (\bibinfo {month}
  {Jul.}\ \bibinfo {year} {2011}),\
  \url{http://adsabs.harvard.edu/abs/2011SSRv..158..289G}%
  \bibAnnoteFile{NoStop}{goossens_resonant_2011}%
\bibitem{priest_solar_1982}%
  \BibitemOpen
  \bibfield{author}{%
  \bibinfo {author} {\bibfnamefont{E.~R.}\ \bibnamefont{Priest}},\ }%
  \emph{\bibinfo {title} {{Solar magneto-hydrodynamics}}},\ Vol.~\bibinfo
  {volume} {23}\ (\bibinfo {year} {1982})\
  \url{http://adsabs.harvard.edu/abs/1982QB539.M23P74}%
  \bibAnnoteFile{NoStop}{priest_solar_1982}%
\bibitem{Berry105}%
  \BibitemOpen
  \bibfield{author}{%
  \bibinfo {author} {\bibfnamefont{M.~V.}\ \bibnamefont{Berry}},\ }%
  in\ \emph{\bibinfo {booktitle} {{Physique des D{\'e}fauts / Physics of
  Defects}}},\ \bibinfo {series} {{Les Houches}}, Vol.\ \bibinfo {volume}
  {XXXV},\ \bibinfo {editor} {edited by\ \bibinfo {editor}
  {\bibfnamefont{R.}~\bibnamefont{Balian}} \emph{et~al.}}\ (\bibinfo {year}
  {1980})\ p.\ \bibinfo {pages} {456}%
  \bibAnnoteFile{NoStop}{Berry105}%
\bibitem{nye_lines_1983}%
  \BibitemOpen
  \bibfield{author}{%
  \bibinfo {author} {\bibfnamefont{J.~F.}\ \bibnamefont{Nye}},\ }%
  \bibfield{journal}{%
  \Doi{10.1098/rspa.1983.0109}{\bibinfo {journal} {Royal Society of London
  Proceedings Series A}}\ }%
  \textbf{\bibinfo {volume} {389}},\ \bibinfo {pages} {279} (\bibinfo {month}
  {Oct.}\ \bibinfo {year} {1983}),\ ISSN \bibinfo {issn} {1364-5021},\
  \url{http://adsabs.harvard.edu/abs/1983RSPSA.389..279N}%
  \bibAnnoteFile{NoStop}{nye_lines_1983}%
\bibitem{edwin_wave_1983}%
  \BibitemOpen
  \bibfield{author}{%
  \bibinfo {author} {\bibfnamefont{P.~M.}\ \bibnamefont{Edwin}}\ and\ \bibinfo
  {author} {\bibfnamefont{B.}~\bibnamefont{Roberts}},\ }%
  \bibfield{journal}{%
  \Doi{10.1007/BF00196186}{\bibinfo {journal} {Solar Physics}}\ }%
  \textbf{\bibinfo {volume} {88}},\ \bibinfo {pages} {179} (\bibinfo {month}
  {Oct.}\ \bibinfo {year} {1983}),\
  \url{http://adsabs.harvard.edu/abs/1983SoPh...88..179E}%
  \bibAnnoteFile{NoStop}{edwin_wave_1983}%
\bibitem{ruderman_nonlinear_2010}%
  \BibitemOpen
  \bibfield{author}{%
  \bibinfo {author} {\bibfnamefont{M.~S.}\ \bibnamefont{Ruderman}}, \bibinfo
  {author} {\bibfnamefont{M.}~\bibnamefont{Goossens}},\ and\ \bibinfo {author}
  {\bibfnamefont{J.}~\bibnamefont{Andries}},\ }%
  \bibfield{journal}{%
  \Doi{10.1063/1.3464464}{\bibinfo {journal} {Physics of Plasmas}}\ }%
  \textbf{\bibinfo {volume} {17}},\ \bibinfo {pages} {2108} (\bibinfo {month}
  {Aug.}\ \bibinfo {year} {2010}),\
  \url{http://adsabs.harvard.edu/abs/2010PhPl...17h2108R}%
  \bibAnnoteFile{NoStop}{ruderman_nonlinear_2010}%
\bibitem{cally_leaky_1986}%
  \BibitemOpen
  \bibfield{author}{%
  \bibinfo {author} {\bibfnamefont{P.~S.}\ \bibnamefont{Cally}},\ }%
  \bibfield{journal}{%
  \Doi{10.1007/BF00147830}{\bibinfo {journal} {Solar Physics}}\ }%
  \textbf{\bibinfo {volume} {103}},\ \bibinfo {pages} {277} (\bibinfo {month}
  {Feb.}\ \bibinfo {year} {1986}),\
  \url{http://adsabs.harvard.edu/abs/1986SoPh..103..277C}%
  \bibAnnoteFile{NoStop}{cally_leaky_1986}%
\bibitem{bogdan_normal_1989}%
  \BibitemOpen
  \bibfield{author}{%
  \bibinfo {author} {\bibfnamefont{T.~J.}\ \bibnamefont{Bogdan}}\ and\ \bibinfo
  {author} {\bibfnamefont{F.}~\bibnamefont{Cattaneo}},\ }%
  \bibfield{journal}{%
  \bibinfo {journal} {The Astrophysical Journal}\ }%
  \textbf{\bibinfo {volume} {342}},\ \bibinfo {pages} {545} (\bibinfo {month}
  {Jul.}\ \bibinfo {year} {1989}),\
  \url{http://adsabs.harvard.edu/abs/1989ApJ...342..545B}%
  \bibAnnoteFile{NoStop}{bogdan_normal_1989}%
\bibitem{centeno_spectropolarimetric_2006}%
  \BibitemOpen
  \bibfield{author}{%
  \bibinfo {author} {\bibfnamefont{R.}~\bibnamefont{Centeno}}, \bibinfo
  {author} {\bibfnamefont{M.}~\bibnamefont{Collados}},\ and\ \bibinfo {author}
  {\bibfnamefont{J.}~\bibnamefont{{Trujillo Bueno}}},\ }%
  \bibfield{journal}{%
  \Doi{10.1086/500185}{\bibinfo {journal} {The Astrophysical Journal}}\ }%
  \textbf{\bibinfo {volume} {640}},\ \bibinfo {pages} {1153} (\bibinfo {month}
  {Apr.}\ \bibinfo {year} {2006}),\ ISSN \bibinfo {issn} {{0004-637X},
  1538-4357},\ \url{http://iopscience.iop.org/0004-637X/640/2/1153}%
  \bibAnnoteFile{NoStop}{centeno_spectropolarimetric_2006}%
\bibitem{allen_orbital_1992}%
  \BibitemOpen
  \bibfield{author}{%
  \bibinfo {author} {\bibfnamefont{L.}~\bibnamefont{Allen}}, \bibinfo {author}
  {\bibfnamefont{M.~W.}\ \bibnamefont{Beijersbergen}}, \bibinfo {author}
  {\bibfnamefont{R.~J.~C.}\ \bibnamefont{Spreeuw}},\ and\ \bibinfo {author}
  {\bibfnamefont{J.~P.}\ \bibnamefont{Woerdman}},\ }%
  \bibfield{journal}{%
  \Doi{10.1103/PhysRevA.45.8185}{\bibinfo {journal} {Physical Review A}}\ }%
  \textbf{\bibinfo {volume} {45}},\ \bibinfo {pages} {8185} (\bibinfo {month}
  {Jun.}\ \bibinfo {year} {1992}),\
  \url{http://link.aps.org/doi/10.1103/PhysRevA.45.8185}%
  \bibAnnoteFile{NoStop}{allen_orbital_1992}%
\bibitem{simpson_optical_1996}%
  \BibitemOpen
  \bibfield{author}{%
  \bibinfo {author} {\bibfnamefont{N.~B.}\ \bibnamefont{Simpson}}, \bibinfo
  {author} {\bibfnamefont{L.}~\bibnamefont{Allen}},\ and\ \bibinfo {author}
  {\bibfnamefont{M.~J.}\ \bibnamefont{Padgett}},\ }%
  \bibfield{journal}{%
  \Doi{10.1080/09500349608230675}{\bibinfo {journal} {Journal of Modern
  Optics}}\ }%
  \textbf{\bibinfo {volume} {43}},\ \bibinfo {pages} {2485} (\bibinfo {year}
  {1996}),\ ISSN \bibinfo {issn} {0950-0340},\
  \url{http://www.tandfonline.com/doi/abs/10.1080/09500349608230675}%
  \bibAnnoteFile{NoStop}{simpson_optical_1996}%
\bibitem{demore_mechanical_2012}%
  \BibitemOpen
  \bibfield{author}{%
  \bibinfo {author} {\bibfnamefont{C.~E.~M.}\ \bibnamefont{Demore}}, \bibinfo
  {author} {\bibfnamefont{Z.}~\bibnamefont{Yang}}, \bibinfo {author}
  {\bibfnamefont{A.}~\bibnamefont{Volovick}}, \bibinfo {author}
  {\bibfnamefont{S.}~\bibnamefont{Cochran}}, \bibinfo {author}
  {\bibfnamefont{M.~P.}\ \bibnamefont{{MacDonald}}},\ and\ \bibinfo {author}
  {\bibfnamefont{G.~C.}\ \bibnamefont{Spalding}},\ }%
  \bibfield{journal}{%
  \Doi{10.1103/PhysRevLett.108.194301}{\bibinfo {journal} {Physical Review
  Letters}}\ }%
  \textbf{\bibinfo {volume} {108}},\ \bibinfo {pages} {194301} (\bibinfo
  {month} {May}\ \bibinfo {year} {2012}),\ ISSN \bibinfo {issn} {0031-9007},\
  \url{http://adsabs.harvard.edu/abs/2012PhRvL.108s4301D}%
  \bibAnnoteFile{NoStop}{demore_mechanical_2012}%
\bibitem{volke-sepulveda_transfer_2008}%
  \BibitemOpen
  \bibfield{author}{%
  \bibinfo {author} {\bibfnamefont{K.}~\bibnamefont{Volke-Sep{\'u}lveda}},
  \bibinfo {author} {\bibfnamefont{A.~O.}\ \bibnamefont{Santill{\'a}n}},\ and\
  \bibinfo {author} {\bibfnamefont{R.~R.}\ \bibnamefont{Boullosa}},\ }%
  \bibfield{journal}{%
  \Doi{10.1103/PhysRevLett.100.024302}{\bibinfo {journal} {Physical Review
  Letters}}\ }%
  \textbf{\bibinfo {volume} {100}},\ \bibinfo {pages} {024302} (\bibinfo {month}
  {Jan.}\ \bibinfo {year} {2008}),\ ISSN \bibinfo {issn} {0031-9007},\
  \url{http://adsabs.harvard.edu/abs/2008PhRvL.100b4302V}%
  \bibAnnoteFile{NoStop}{volke-sepulveda_transfer_2008}%
\end{thebibliography}
 %

% \begin{figure}
% \caption{}
% \label{}
% \end{figure}

% \begin{table}
% \caption{}
% \label{}
% \end{tabular}
% \end{table}

\end{document}